\pdfoutput=1
\documentclass[letterpaper,10pt,conference]{ieeeconf}
\usepackage{amssymb}
\usepackage{amsmath}
\usepackage{graphicx}
\usepackage[thinlines]{easybmat}
\usepackage{booktabs}
\usepackage{xcolor}

\IEEEoverridecommandlockouts    
\newcommand{\E}{\operatorname{E}}

\newcommand{\MSE}{\operatorname{MSE}}

\newcommand{\N}{\mathcal{N}}

\newcommand{\bigo}{\mathcal{O}}

\renewcommand{\t}{\mathrm{T}}

\renewcommand{\hbar}{\bar{h}}
\newcommand{\p}{\mathrm{p}}
\renewcommand{\v}{\mathrm{v}}

\newcommand{\yb}{\mathbf{y}{}}

\newcommand{\xb}{\mathbf{x}{}}

\newcommand{\bb}{\mathbf{b}{}}

\newcommand{\vb}{\mathbf{v}{}}
\newcommand{\wb}{\mathbf{w}{}}
\newcommand{\Qb}{\mathbf{Q}{}}
\newcommand{\Rb}{\mathbf{R}{}}
\newcommand{\Sb}{\mathbf{S}{}}
\newcommand{\Ib}{\mathbf{I}{}}
\newcommand{\zerob}{\mathbf{0}{}}
\newcommand{\mub}{\boldsymbol{\mu}{}}
\newcommand{\Sigmab}{\boldsymbol{\Sigma}{}}

\newcommand{\Hb}{\mathbf{H}{}}
\newcommand{\Hbbar}{\mathbf{\overline{H}}{}}

\newcommand{\Kb}{\mathbf{K}{}}
\newcommand{\Lb}{\mathbf{L}{}}
\newcommand{\Cb}{\mathbf{C}{}}
\newcommand{\Pb}{\mathbf{P}{}}
\newcommand{\Fb}{\mathbf{F}{}}

\newcommand{\eb}{\mathbf{e}{}}

\newcommand{\xbhat}{\mathbf{\hat{x}}{}}
\newcommand{\xbbar}{\mathbf{\bar{x}}{}}
\newcommand{\ybbar}{\mathbf{\bar{y}}{}}
\newcommand{\vbbar}{\mathbf{\bar{v}}{}}
\newcommand{\Rbbar}{\mathbf{\overline{R}}{}}

\newcommand{\vbtilde}{\mathbf{\widetilde{v}}{}}
\newcommand{\wbtilde}{\mathbf{\widetilde{w}}{}}

\title{Performance of the Kalman Filter and Smoother for Benchmark Studies}

\author{Bat\i n Kurt$^{1}$ and Umut Orguner$^{1}$
\thanks{$^{1}$Department of Electrical and Electronics Engineering, 
Middle East Technical University, 06800 Ankara, Turkey.
(e-mail:\{batin.kurt, umut\}@metu.edu.tr).}
\thanks{*Bat\i n Kurt is supported by Turk Telekom within the framework of the 5G and Beyond Joint Graduate Support Programme coordinated by the Information and Communication Technologies Authority of Turkey, and by TUBITAK (The Scientific and Technological Research Council of Turkey) under the 2224-A Grant Program for Participation in Scientific Events Abroad.}
\thanks{This work has been submitted to the IEEE for possible publication.
Copyright may be transferred without notice, after which this version may no longer be accessible.}
}
\begin{document}

\maketitle

\begin{abstract} 
We propose analytical mean square error (MSE) expressions for the Kalman filter (KF) and the Kalman smoother (KS) for benchmark studies, where the true system dynamics are unknown or unavailable to the estimator. In such cases, as in benchmark evaluations for target tracking, the analysis relies on deterministic state trajectories. This setting introduces a model mismatch between the estimator and the true system, causing the covariance estimates to no longer reflect the actual estimation errors. To enable accurate performance prediction for deterministic state trajectories without relying on computationally intensive Monte Carlo simulations, we derive recursive MSE expressions with linear time complexity. The proposed framework also accounts for measurement model mismatch and provides an efficient tool for performance evaluation in benchmark studies involving long trajectories. Simulation results confirm the accuracy and computational efficiency of the proposed method.
\end{abstract}

\section{Introduction}\label{sec:introduction}
State estimation is the process of inferring the hidden states of a dynamical system, denoted by $\xb_k$, from noisy and uncertain measurements $\yb_k$, and it serves as a cornerstone of modern control~\cite{simon2006optimal,stengel1994optimal}, robotics~\cite{Thrun2005Robotics,Barfoot2024},  signal processing~\cite{Kay_EstimationTheory,Haykin2002Adaptive}, and navigation~\cite{Groves2013}. A fundamental solution for state estimation is the Kalman filter (KF), introduced by R.~E.~Kalman in 1960~\cite{Kalman1960}. The Kalman filter provides the minimum mean square error (MMSE) estimates of the system state, shown as $\xbhat_{k|k}$, when the system of interest is linear and Gaussian. Owing to its recursive structure, the KF can process measurements sequentially as they become available, making it well suited for real-time applications. For offline estimation problems that allow access to measurements beyond the current time, smoothing algorithms such as the Kalman smoother (KS) can be used to refine the state estimates.
The KS yields the MMSE-optimal estimates of the system states, $\xbhat_{k|K}$, when all measurements within the estimation interval $k = 0, \ldots, K$ are available. The KS has several equivalent implementations that differ in computational structure but yield identical MMSE estimates.
Notable examples include the Rauch–Tung–Striebel (RTS) smoother~\cite{RTS1965} and the Mayne–Fraser two-filter smoother~\cite{MAYNE1966,Fraser1967}.

In the standard linear–Gaussian setting, the KF and KS are Bayesian estimation algorithms that infer the posterior distribution of the state given the available measurements, denoted $p(\xb_k |\yb_{0:k})$ for filtering and $p(\xb_k | \yb_{0:K})$ for smoothing. When the assumed state and measurement models coincide with the true dynamical system (i.e., there is no model mismatch), these posteriors are Gaussian and are fully characterized by their means and covariances, $(\xbhat_{k|k},\, \Pb_{k|k})$ and $(\xbhat_{k|K},\, \Pb_{k|K})$, respectively. Under these conditions, the Kalman filter and smoother are unbiased estimators in the Bayesian sense~\cite[Section~2.5]{bar2001estimation}:
\begin{align}
\E\big[\xbhat_{k|k}(\yb_{0:k})\big] &= \E[\xb_k], &
\E\big[\xbhat_{k|K}(\yb_{0:K})\big] &= \E[\xb_k],
\label{eqn:KFKSunbiasedness}
\end{align}
where the expectations are taken with respect to the state vector $\xb_k$ and the corresponding measurement sequence $\yb_{0:k}$ for the KF and $\yb_{0:K}$ for the KS.

In the absence of model mismatch, the posterior covariance matrices $\Pb_{k|k}$ and $\Pb_{k|K}$ coincide with the corresponding MSE matrices, that is,
\begin{subequations}
\label{eqn:KFKSconsistency}
\begin{align}
\E\big[\big(\xb_k-\xbhat_{k|k}(\yb_{0:k})\big)\big(\xb_k-\xbhat_{k|k}(\yb_{0:k})\big)^\t\big]=&\,\Pb_{k|k},\\ 
\E[\big(\xb_k-\xbhat_{k|K}(\yb_{0:K})\big)\big(\xb_k-\xbhat_{k|K}(\yb_{0:K})\big)^\t\big]=&\,\Pb_{k|K}.
\end{align}
\end{subequations}
This property is particularly useful for assessing the estimation performance of the KF and KS, as it avoids the need for Monte Carlo simulations to compute the MSE.
The posterior covariance matrices $\Pb_{k|k}$ and $\Pb_{k|K}$ directly represent the true MSE characteristics of the estimators, reflecting the consistency of the KF and the KS~\cite[Section 5.4]{bar2001estimation}.
However, real-life problems often include model mismatch, where the assumed state or measurement models used by the estimator deviate from the true underlying system dynamics.

One particularly relevant form of model mismatch occurs in benchmark studies, where the true system dynamics are unknown or unavailable to the estimator.
In such scenarios, the corresponding state trajectory is typically provided as data, either from simulations or from field experiments conducted during real-world operations.
Since the governing model is not explicitly known, the available state trajectory $\xbbar_k$, $k = 0, \ldots, K$, can be represented as a deterministic sequence, denoted by $\xbbar_{0:K}$.
For instance, in target tracking applications, benchmark datasets commonly use fixed, predefined state trajectories to assess the estimation performance of different algorithms~\cite{BlairWKB1998,BlackmanDBP1999}.

In benchmark studies where the state trajectory is fixed and predefined, multiple realizations of the measurement sequence $\yb_{0:K}$ are generated randomly from the fixed state trajectory $\xbbar_{0:K}$. As a result, a form of model mismatch arises: the true system follows the fixed trajectory $\xbbar_{0:K}$, while the Kalman filter and smoother operate under an assumed probabilistic state-space model that governs the state evolution. Under this mismatch, the properties of unbiasedness and consistency in~\eqref{eqn:KFKSunbiasedness} and~\eqref{eqn:KFKSconsistency} are no longer valid. Instead, the KF/KS estimates become biased and inconsistent (in a non-random (or frequentist) sense)~\cite[Section 2.3]{Kay_EstimationTheory}:
\begin{align} \label{eqn:KFKSbiasedness} \E\big[\xbhat_{k|k}(\yb_{0:k}) \big| \xbbar_{0:k}\big] &\neq \xbbar_k, &
\E\big[\xbhat_{k|K}(\yb_{0:K}) \big| \xbbar_{0:K}\big] &\neq \xbbar_k,
\end{align} and
\begin{subequations} \label{eqn:KFKSinconsistency} \begin{align} \E\big[\big(\xb_k - \xbhat_{k|k}(\yb_{0:k})\big)\big(\xb_k - \xbhat_{k|k}(\yb_{0:k})\big)^\t \big| \xbbar_{0:k}\big] &\neq \Pb_{k|k},\\ \E\big[\big(\xb_k - \xbhat_{k|K}(\yb_{0:K})\big)\big(\xb_k - \xbhat_{k|K}(\yb_{0:K})\big)^\t \big| \xbbar_{0:K}\big] &\neq \Pb_{k|K}, \end{align} \end{subequations} 
where the expectations are taken only over the random measurement sequences, $\yb_{0:k}$ for the KF and $\yb_{0:K}$ for the KS, since the state sequence is deterministic and fixed as $\xb_{0:K} = \xbbar_{0:K}$.

Due to this model mismatch, the covariance estimates $\Pb_{k|k}$ and $\Pb_{k|K}$ no longer represent the actual estimation errors. Consequently, accurately predicting the estimation performance of the Kalman filter and smoother becomes a nontrivial task. This challenge motivates the development of efficient analytical methods that can predict the true estimation performance for the KF and the KS under model mismatch without relying on computationally costly Monte Carlo simulations. To the best of the authors’ knowledge, this topic has received limited attention in the literature, with existing studies mainly addressing performance bounds~\cite{Fritsche2016}, measurement model misspecification~\cite{SangsukB1990,GeSDW2016}, or both~\cite{TeichnerM2023KF,TeichnerM2023KS}. However, the majority of existing methods are not exact, offering only performance bounds, or they treat only partial cases of model misspecification, such as inaccuracies in the measurement model. 

Motivated by these limitations, in a previous study, we presented batch and horizon-recursive analytical MSE expressions for fixed state trajectories, including cases with measurement model mismatch~\cite{KurtOrguner2025}. While the methods in~\cite{KurtOrguner2025} provide exact analytical results, the computational complexity of MSE expressions for smoothing grew quadratically with trajectory length. In this paper, we propose computationally efficient MSE recursions for benchmark scenarios with fixed state trajectories. The proposed method retains the generality of the previous formulation and remains applicable under measurement model mismatch, while reducing the computational complexity to linear in the trajectory length $K$, i.e., $\bigo(K\,n_x^3)$, where $n_x$ denotes the state dimension. The validity of the proposed approach is demonstrated through simulation studies. As a result, the method enables accurate performance prediction for long trajectories with significantly reduced computational cost.

The rest of the paper is organized as follows: Section~\ref{sec:problem_definition} defines the problem. Section~\ref{sec:modified_two_filter_KS} introduces the modified two-filter formulation of the KS~\cite{WallWS1981}, which serves as the foundation for our proposed method. In Section~\ref{sec:AnalyticalRecursiveMSE},
analytical recursive expressions for the MSE are obtained. Section~\ref{sec:simulation_results} presents simulation results that demonstrate the accuracy and efficiency of the proposed method. Section~\ref{sec:conclusions} concludes the paper.


\section{Problem Definition}\label{sec:problem_definition}
We consider a system where the true state evolution model is unknown but the corresponding state trajectory $\xbbar_{0:K}$ is available as data.
As the governing model is unknown, the available state trajectory $\xbbar_k \in \mathbb{R}^{n_x}$, $k = 0, \ldots, K$, is represented as a deterministic sequence, denoted by $\xbbar_{0:K}$. 
In this study, our objective is to predict the mean squared error (MSE) performance of the Kalman filter and smoother in estimating the state trajectory $\xbbar_{0:K}$ from the measurements $\yb_k \in \mathbb{R}^{n_y}$, $k = 0, \ldots, K$, which are related to $\xbbar_k$ by
\begin{align}\label{eqn:true_measurement_model}
    \yb_k = \,\Hbbar\xbbar_k + \vbbar_k,
\end{align}
for $k=0,\ldots,K$, where
\begin{itemize}
    \item $\Hbbar \in \mathbb{R}^{n_y\times n_x}$ is the true measurement matrix\footnote{Extension to nonlinear measurement functions is straightforward and omitted for simplicity.};
    \item $\vbbar_k\sim p_{\vbbar}(\vbbar_k)$ is the true zero-mean white measurement noise with true covariance $\Rbbar$. Note that $p_{\vbbar}(\cdot)$ is an arbitrary probability density function that is not necessarily Gaussian.
\end{itemize}
Based on the true measurement model in \eqref{eqn:true_measurement_model},
the conditional distribution of the measurement sequence $\yb_{0:K}$ given the true state sequence $\xbbar_{0:K}$, i.e., the true likelihood function $\bar{p}(\yb_{0:K}|\xbbar_{0:K})$, can be expressed as
\begin{align}
\bar{p}(\yb_{0:K}|\xbbar_{0:K})=\prod_{k=0}^K p_{\vbbar}(\yb_k-\Hbbar\xbbar_k).
\end{align}
The Kalman filter and smoother do not have access to the unknown state evolution model that generates the true state trajectory $\xbbar_{0:K}$, nor to the trajectory itself. They instead assume a mismatched state model of the form
\begin{align}\label{eqn:assumed_state_equation}
        \xb_{k+1}=&\, \Fb \xb_k+\wb_k,
\end{align}
where 
\begin{itemize}
\item $\xb_{k}\in\mathbb{R}^{n_x}$, $k=0,\ldots,K$, is the state sequence to be estimated with the initial state $\xb_0$ assumed to be distributed as $\xb_0\sim\N(\xb_0;\mub_0,\Sigmab_0)$;
\item $\Fb \in \mathbb{R}^{n_x\times n_x}$ is the assumed state matrix;
\item $\wb_k\sim\N(\wb_k; \zerob,\Qb)$ is the assumed additive white Gaussian process noise.
\end{itemize}
Similarly, the Kalman filter and smoother assume a (possibly mismatched) measurement model of the form 
\begin{align}\label{eqn:assumed_measurement_model}
        \yb_k=&\,\Hb \xb_k+\vb_k, 
\end{align}
for $k=0,\ldots,K$, where
\begin{itemize}
\item $\Hb \in \mathbb{R}^{n_y\times n_x}$ is the assumed measurement matrix;
\item $\vb_k\sim\N(\vb_k;\zerob,\Rb)$ is the assumed additive white Gaussian measurement noise.
\end{itemize} 
Note that the noisy measurements $\yb_k$ are generated according to the true measurement model in \eqref{eqn:true_measurement_model}, whereas the Kalman filter and smoother assume the (possibly mismatched) model in \eqref{eqn:assumed_measurement_model}.

The Kalman filter calculates the estimate $\xbhat_{k|k}$ with the covariance $\Pb_{k|k}$ for the true state $\xbbar_k$ using the following forward recursion. 
\begin{subequations}\label{eqn:KalmanFilter}
    \begin{align}
            \xbhat_{k|k} =& \,\xbhat_{k|k-1} + \Kb_k\left(\yb_k-\Hb\xbhat_{k|k-1}\right), \\
            \Pb_{k|k} =& \, \Pb_{k|k-1} - \Kb_k\Sb_k\Kb_k^\t, 
    \end{align}
\end{subequations}
for $k=0,\ldots,K$, where
\begin{subequations}
\begin{align}
\xbhat_{k|k-1}=\,&\begin{cases}\mub_0,&k=0\\\Fb\xbhat_{k-1|k-1},&\text{otherwise}\end{cases},\\
\Pb_{k|k-1}=&\,\begin{cases}\Sigmab_0,&k=0\\\Fb\Pb_{k-1|k-1}\Fb^\t+\Qb,&\text{otherwise}\end{cases},\\
\Sb_k =&\, \Hb\Pb_{k|k-1}\Hb^\t+\Rb,\\
\Kb_k =&\, \Pb_{k|k-1}\Hb^\t \Sb_k^{-1}.
\end{align}
\end{subequations}
Note that the Kalman filter estimate $\xbhat_{k|k}$ is a function of $\yb_{0:k}$, i.e., $\xbhat_{k|k}=\xbhat_{k|k}(\yb_{0:k})$.

Similarly, the Kalman smoother calculates the estimate $\xbhat_{k|K}$ with the covariance $\Pb_{k|K}$.
Among the various algorithms available to obtain the smoothed estimates $\xbhat_{k|K}$,
the RTS smoother computes the estimates $\xbhat_{k|K}$ and covariances $\Pb_{k|K}$ using the following backward recursion.
\begin{subequations}\label{eqn:KalmanSmoother}
    \begin{align}
            \xbhat_{k|K} =&\,\xbhat_{k|k} + \Lb_k\left(\xbhat_{k+1|K}-\xbhat_{k+1|k}\right), \\
            \Pb_{k|K} = & \,\Pb_{k|k} + \Lb_k\left(\Pb_{k+1|K}-\Pb_{k+1|k}\right)\Lb_k^\t,
    \end{align}
\end{subequations}
for $k=K-1,\ldots,0$, where 
\begin{align}
    \Lb_k =&\, \Pb_{k|k}\Fb^\t \Pb_{k+1|k}^{-1}.
\end{align}
Note that the Kalman smoother estimate $\xbhat_{k|K}$ is a function of $\yb_{0:K}$, i.e., $\xbhat_{k|K}=\xbhat_{k|K}(\yb_{0:K})$.

The objective of this study is to derive the mean square error (MSE) matrices $\MSE_{k|K} \in \mathbb{R}^{n_x \times n_x}$ and $\MSE_{k|k} \in \mathbb{R}^{n_x \times n_x}$ corresponding to the Kalman smoother and filter estimates, respectively, which are defined as follows.
\begin{subequations}\label{eqn:MSEdefinitionForFilterandSmoother}
\begin{align}
    \MSE_{k|K}\triangleq&\,\E_{\bar{p}}\big[\big(\xbhat_{k|K}(\yb_{0:K})-\xbbar_k\big)\big(\cdot\big)^\t\big|\xbbar_{0:K}\big],\\
    \MSE_{k|k}\triangleq&\,\E_{\bar{p}}\big[\big(\xbhat_{k|k}(\yb_{0:k})-\xbbar_k\big)\big(\cdot\big)^\t\big|\xbbar_{0:k}\big],
\end{align}
\end{subequations}
for $k=0,\ldots,K$, where the expectations $\E_{\bar{p}}[\cdot]$ are taken with respect to the true likelihood functions $\bar{p}(\yb_{0:K}|\xbbar_{0:K})$ for the Kalman smoother and $\bar{p}(\yb_{0:k}|\xbbar_{0:k})$ for the Kalman filter. The notation $(\mathbf{a})(\cdot)^\t$ denotes the outer product $(\mathbf{a})(\mathbf{a})^\t$ for brevity.

To derive recursions for the MSE matrices in \eqref{eqn:MSEdefinitionForFilterandSmoother} with linear time complexity, we adopt the modified two-filter Kalman smoother proposed by Wall et al.~\cite{WallWS1981}, given in Section~\ref{sec:modified_two_filter_KS}. This formulation extends the original two-filter smoothing approach introduced by Mayne and Fraser~\cite{MAYNE1966,Fraser1967}.
\section{Modified Two-Filter Formulation of the Kalman Smoother}\label{sec:modified_two_filter_KS}
Consider the reversed-time process $\xb_{0:K}^b$, which is sample-path equivalent to the forward-time process $\xb_{0:K}$ for a given prior $\xb_0 \sim \mathcal{N}(\xb_0;\mub_0, \Sigmab_0)$.

The prior
$\N(\xb_0;\mub_0,\Sigmab_0)$ induces the marginal distributions
$\xb_k\sim\N(\xb_k;\mub_k,\Sigmab_k)$ for $k=0,\ldots,K$, where the means $\mub_{0:K}$ and the covariances $\Sigmab_{0:K}$ can be computed recursively as follows.
\begin{subequations}
\begin{align}
\mub_k=&\,\begin{cases}\mub_0,&k=0\\ \Fb\mub_{k-1},&k>0\end{cases}, \\
\Sigmab_k=&\,\begin{cases}\Sigmab_0,&k=0\\\Fb\Sigmab_{k-1}\Fb^\t+\Qb,&k>0\end{cases}.
\end{align}
\end{subequations}
The reversed-time process can now be defined as
\begin{align}
\xb_k^b=&\,\Fb_{k+1}^b\xb_{k+1}^b+\wb_{k+1}^b,
\end{align}
for $k=K-1,\ldots,0$, with $\xb_K^b\sim\N(\xb_K^b;\mub_K,\Sigmab_K)$ and $\wb_{k+1}^b\sim\N(\wb_{k+1}^b;\mub_k-\Fb_k^b\mub_{k+1},\Qb_{k+1}^b)$ where
\begin{subequations}
\begin{align}
\Fb_{k+1}^b\triangleq&\,\Sigmab_k\Fb^\t\Sigmab_{k+1}^{-1},\\
\Qb_{k+1}^b\triangleq&\,\Sigmab_k-\Sigmab_k\Fb^\t\Sigmab_{k+1}^{-1}\Fb\Sigmab_k.
\end{align}
\end{subequations}
It follows that $\xb_k^b\sim\N(\xb_k^b;\mub_k,\Sigmab_k)$ for $k=0,\ldots,K$.

The reversed-time Kalman filter is then defined to compute the estimate $\xbhat_{k|k:K}$ with covariance $\Pb_{k|k:K}$ through the following backward recursion.
\begin{subequations}\label{eqn:BackwardKalmanFilter}
    \begin{align}
            \xbhat_{k|k:K} = &\,\xbhat_{k|k+1:K} + \Kb_{k|k+1:K}\big(\yb_k-\Hb\xbhat_{k|k+1:K}\big), \\
            \Pb_{k|k:K} = &\,\Pb_{k|k+1:K} - \Kb_{k|k+1:K}\Sb_{k|k+1:K}\Kb_{k|k+1:K}^\t, 
    \end{align}
\end{subequations}
for $k=K,\ldots,0$, where
\begin{subequations}
\begin{align}
\xbhat_{k|k+1:K}=&\,\begin{cases}\mub_K,&k=K\\
\mub_k+\Fb_{k+1}^b(\xbhat_{k+1|k+1:K}-\mub_{k+1}),&k<K\end{cases},\\
\Pb_{k|k+1:K}=&\,\begin{cases}\Sigmab_K,&k=K\\\Fb_{k+1}^b\Pb_{k+1|k+1:K}\Fb_{k+1}^{b,\t}+\Qb_{k+1}^b,&k<K\end{cases},\\
\Sb_{k|k+1:K} =&\, \Hb\Pb_{k|k+1:K}\Hb^\t+\Rb,\\
\Kb_{k|k+1:K} =&\, \Pb_{k|k+1:K}\Hb^\t\Sb_{k|k+1:K}^{-1}.
\end{align}
\end{subequations}
Note that the reversed-time Kalman filter estimate $\xbhat_{k|k:K}$ is a function of the measurements $\yb_{k:K}$, i.e., $\xbhat_{k|k:K}=\xbhat_{k|k:K}(\yb_{k:K})$.

The modified two-filter Kalman smoother computes the estimates $\xbhat_{k|0:K}$ and covariances $\Pb_{k|0:K}$, given by
\begin{subequations}
\begin{align}
\xbhat_{k|0:K}=&\,\Pb_{k|0:K}\Big(\Pb_{k|0:k}^{-1}\xbhat_{k|0:k}+\Pb_{k|k+1:K}^{-1}\xbhat_{k|k+1:K} \nonumber \\ & \,-\Sigmab_k^{-1}\mub_k\Big),\\
\Pb_{k|0:K}=&\,\Big(\Pb_{k|0:k}^{-1}+\Pb_{k|k+1:K}^{-1}-\Sigmab_k^{-1}\Big)^{-1},
\end{align}
\end{subequations}
for $k=0,\ldots,K$.
\section{MSE Recursions with Linear Time Complexity}\label{sec:AnalyticalRecursiveMSE}
The MSE matrix $\MSE_{k|\ell:m}$ corresponding to the estimate $\xbhat_{k|\ell:m}$ is defined as
\begin{align}\label{eqn:MSEdefinition}
    \MSE_{k|\ell:m}\triangleq&\,\E_{\bar{p}}\big[\big(\xbhat_{k|\ell:m}(\yb_{\ell:m})-\xbbar_k\big)\big(\cdot\big)^\t{}\big|\xbbar_{0:K}\big],
\end{align}
where the expectation is taken with respect to $\bar{p}(\yb_{\ell:m}|\xbbar_{0:K})$.
This matrix can be decomposed as follows:
\begin{align}\label{eqn:MSE_k_ell_m}
\MSE_{k|\ell:m}=\,
\Cb_{k|\ell:m}+\bb_{k|\ell:m}\bb_{k|\ell:m}^\t,
\end{align}
where
\begin{subequations}
\begin{align}
\Cb_{k|\ell:m}\triangleq& \,\E_{\bar{p}}[\eb_{k|\ell:m}\eb_{k|\ell:m}^\t\big|\xbbar_{0:K}],\\
\eb_{k|\ell:m}\triangleq&\,\xbhat_{k|\ell:m}(\yb_{\ell:m})-\xbhat_{k|\ell:m}^*,\\
\bb_{k|\ell:m}\triangleq&\, \xbhat_{k|\ell:m}^*-\xbbar_{k},\\
\xbhat_{k|\ell:m}^*\triangleq&\,\E_{\bar{p}}[\xbhat_{k|\ell:m}(\yb_{\ell:m})|\xbbar_{0:K}] = \,\xbhat_{k|\ell:m}(\ybbar_{\ell:m}), \\
\ybbar_k \triangleq& \,\Hbbar \xbbar_k,
\end{align}
\end{subequations}
for $k = 0, \ldots, K$.
The quantity $\xbhat_{k|\ell:m}^*$ corresponds to the estimate obtained under noiseless measurements. The decomposition in \eqref{eqn:MSE_k_ell_m} allows the independent evaluation of the contributions from measurement noise, represented by the covariance term $\Cb_{k|\ell:m}$, and from model mismatch in the state equation, represented by the squared bias term  $\bb_{k|\ell:m}\bb_{k|\ell:m}^\t$.

To derive recursive expressions for the MSE matrices in \eqref{eqn:MSEdefinitionForFilterandSmoother}, we first obtain recursions for the intermediate quantities $\xbhat_{k|0:k}^*$, $\xbhat_{k|0:K}^*$, $\eb_{k|0:k}$, and $\eb_{k|0:K}$, which are then used to derive the recursions for the sufficient statistics 
$\bb_{k|0:k}$, $\bb_{k|0:K}$, $\Cb_{k|0:k}$, and $\Cb_{k|0:K}$. Note that $(\cdot)_{k|0:k}=(\cdot)_{k|k}$ and $(\cdot)_{k|0:K}=(\cdot)_{k|K}$.

The derivation employs the modified two-filter formulation of the Kalman smoother, described in detail in Section~\ref{sec:modified_two_filter_KS}.

\subsection{Recursions for $\xbhat_{k|0:k}^*$ and $\xbhat_{k|0:K}^*$}
The forward estimate $\xbhat_{k|0:k}^*$ can be obtained recursively as
\begin{align}
            \xbhat_{k|0:k}^* =&\, \xbhat_{k|0:k-1}^* + \Kb_{k|0:k-1}\big(\ybbar_k-\Hb\xbhat_{k|0:k-1}^*\big), 
    \end{align}
for $k=0,\ldots,K$, where
\begin{align}
\xbhat_{k|0:k-1}^*=&\,\begin{cases}\mub_0,&k=0\\ \Fb\xbhat_{k-1|0:k-1}^*,&k>0\end{cases}.
\end{align}
The reversed-time estimate $\xbhat_{k|k:K}^*$ can be computed as
    \begin{align}
            \xbhat_{k|k:K}^* = &\,\xbhat_{k|k+1:K}^* + \Kb_{k|k+1:K}\big(\ybbar_k-\Hb\xbhat_{k|k+1:K}^*\big),    \end{align}
for $k=K,\ldots,0$, where
\begin{align}
\hspace{-0.5cm}\xbhat_{k|k+1:K}^*=\,\begin{cases}\mub_K,&k=K\\\mub_k+\Fb_{k+1}^b(\xbhat_{k+1|k+1:K}^*-\mub_{k+1}),&k<K\end{cases}.
\end{align}
The forward and reversed-time estimates can then be combined to obtain the smoothed estimate $\xbhat_{k|0:K}^*$ as
\begin{align}
\xbhat_{k|0:K}^*=&\,\Pb_{k|0:K}\Big(\Pb_{k|0:k}^{-1}\xbhat_{k|0:k}^*\nonumber \\ &\,+\Pb_{k|k+1:K}^{-1}\xbhat_{k|k+1:K}^*-\Sigmab_k^{-1}\mub_k\Big).
\end{align}

\subsection{Recursions for $\eb_{k|0:k}$ and $\eb_{k|0:K}$}
The forward error term $\eb_{k|0:k}$ is given by
\begin{align}
         \eb_{k|0:k} =&\, \eb_{k|0:k-1} + \Kb_{k|0:k-1}\big(\vbbar_k-\Hb\eb_{k|0:k-1}\big)
\end{align}
for $k=0,\ldots,K$, where
\begin{align}
\eb_{k|0:k-1}=&\,\begin{cases}\zerob,&k=0\\\Fb\eb_{k-1|0:k-1},&k>0\end{cases}.
\end{align}
Similarly, the reversed-time error term $\eb_{k|k:K}$ is given by
    \begin{align}
            \eb_{k|k:K} =&\, \eb_{k|k+1:K}  +  \Kb_{k|k+1:K}\big(\vbbar_k-\Hb\eb_{k|k+1:K}\big),
    \end{align}
for $k=K,\ldots,0$, where
\begin{align}
\eb_{k|k+1:K}=&\,\begin{cases}\zerob,&k=K\\\Fb_{k+1}^b\eb_{k+1|k+1:K},&
k<K\end{cases}.
\end{align}
The forward and reversed-time error terms are then combined to calculate the smoothed error term $\eb_{k|0:K}$ as
\begin{align}
\eb_{k|0:K}=&\, \Pb_{k|0:K}\Big(\Pb_{k|0:k}^{-1}\eb_{k|0:k}+\Pb_{k|k+1:K}^{-1}\eb_{k|k+1:K}\Big).
\end{align}

\subsection{Recursions for $\bb_{k|0:k}$ and $\bb_{k|0:K}$}
Using the forward recursions for $\xbhat_{k|0:k}^*$, the forward bias term 
$\bb_{k|0:k}$ can be obtained recursively as
\begin{align}\label{eqn:forward_bias_eqn}
            \bb_{k|0:k}=&\,\bb_{k|0:k-1}+\Kb_{k|0:k-1}(\vbtilde_k-\Hb\bb_{k|0:k-1}),
    \end{align}
for $k=0,\ldots,K$, where
\begin{subequations}    
\begin{align}
\bb_{k|0:k-1}=&\,\begin{cases}\bb_0,&k=0\\ \Fb\bb_{k-1|0:k-1}-\wbtilde_k,&k>0\end{cases},\\
\bb_0\triangleq&\, \mub_0-\xbbar_0,\\
\wbtilde_k\triangleq&\,\xbbar_k-\Fb\xbbar_{k-1},\\
\vbtilde_k\triangleq&\,\left(\Hbbar-\Hb\right)\xbbar_k.
\end{align}
\end{subequations}
Using the reversed-time recursions for $\xbhat_{k|k:K}^*$, the reversed-time bias term $\bb_{k|k:K}$ can be obtained recursively as
    \begin{align}
            \bb_{k|k:K} = &\,\bb_{k|k+1:K} + \Kb_{k|k+1:K}\big(\vbtilde_k-\Hb\bb_{k|k+1:K}\big), 
    \end{align}
for $k=K,\ldots,0$, where
\begin{subequations}
\begin{align}
\hspace{-0.5cm}\bb_{k|k+1:K}=&\,\begin{cases}\bb_K,&k=K\\ \bb_k+\Fb_{k+1}^b\big(\bb_{k+1|k+1:K}-\bb_{k+1}\big),&k<K\end{cases},\\
\bb_k\triangleq&\,\mub_k-\xbbar_k.
\end{align}
\end{subequations}
The forward and reversed-time bias terms can then be combined to compute the smoothed bias term $\bb_{k|0:K}$ as
\begin{align}\label{eqn:combined_bias_eqn}
\bb_{k|0:K}=&\,\Pb_{k|0:K}\big(\Pb_{k|0:k}^{-1}\bb_{k|0:k}\nonumber \\&\,+\Pb_{k|k+1:K}^{-1}\bb_{k|k+1:K}-\Sigmab_k^{-1}\bb_k\big).
\end{align}

\subsection{Recursions for $\Cb_{k|0:k}$ and $\Cb_{k|0:K}$}

Using the forward recursions for the error term $\eb_{k|0:k}$, the forward covariance term $\Cb_{k|0:k}$ is given by
\begin{align}\label{eqn:forward_covariance_eqn}
         \Cb_{k|0:k} = &\,\left(\Ib-\Kb_{k|0:k-1}\Hb\right)\Cb_{k|0:k-1}
         \left(\Ib-\Kb_{k|0:k-1}\Hb\right)^\t \nonumber \\ &\,+ \Kb_{k|0:k-1}\Rbbar\Kb_{k|0:k-1}^\t,
\end{align}
for $k=0,\ldots,K$, where
\begin{align}
\Cb_{k|0:k-1}=&\,\begin{cases}\zerob,&k=0\\\Fb\Cb_{k-1|0:k-1}\Fb^\t,&k>0\end{cases}.
\end{align}
Similarly, using the reversed-time recursions for the error term $\eb_{k|k:K}$, the reversed-time covariance term $\Cb_{k|k:K}$ is given by
    \begin{align}
            \Cb_{k|k:K} = &\,\left(\Ib-\Kb_{k|k+1:K}\Hb\right)\Cb_{k|k+1:K}
            \left(\Ib-\Kb_{k|k+1:K}\Hb\right)^\t \nonumber \\ &\,+ \Kb_{k|k+1:K}\Rbbar\Kb_{k|k+1:K}^\t,         
    \end{align}
for $k=K,\ldots,0$, where
\begin{align}
\Cb_{k|k+1:K}=&\,\begin{cases}\zerob,&k=K\\\Fb_{k+1}^b\Cb_{k+1|k+1:K}\Fb_{k+1}^{b,\t},&k<K\end{cases}.
\end{align}
The forward and reversed-time covariance terms are then combined to calculate the smoothed covariance term $\Cb_{k|0:K}$ as
\begin{align}\label{eqn:combined_covariance_eqn}
\Cb_{k|0:K}=&\,\Pb_{k|0:K}\big(\Pb_{k|0:k}^{-1}\Cb_{k|0:k}\Pb_{k|0:k}^{-1}\nonumber \\ &\,+\Pb_{k|k+1:K}^{-1}\Cb_{k|k+1:K}\Pb_{k|k+1:K}^{-1}\big)\Pb_{k|0:K}.
\end{align}
\subsection{Expressions for $\MSE_{k|k}$ and $\MSE_{k|K}$}
The MSE matrices defined in \eqref{eqn:MSEdefinitionForFilterandSmoother} are given as
\begin{subequations}
    \begin{align}
    \MSE_{k|k} = \Cb_{k|0:k} + \bb_{k|0:k}\bb_{k|0:k}^\t, \\
    \MSE_{k|K} = \Cb_{k|0:K} + \bb_{k|0:K}\bb_{k|0:K}^\t,
\end{align}
\end{subequations}
where sufficient statistics $\bb_{k|0:k}$, $\bb_{k|0:K}$, $\Cb_{k|0:k}$, and $\Cb_{k|0:K}$ can be obtained recursively using their expressions
in \eqref{eqn:forward_bias_eqn}-\eqref{eqn:combined_bias_eqn} and \eqref{eqn:forward_covariance_eqn}-\eqref{eqn:combined_covariance_eqn}.

\section{Simulation Results}\label{sec:simulation_results}
\begin{figure}[!t]
\vspace{5pt}
  \centering
  \includegraphics[width=0.6\linewidth]{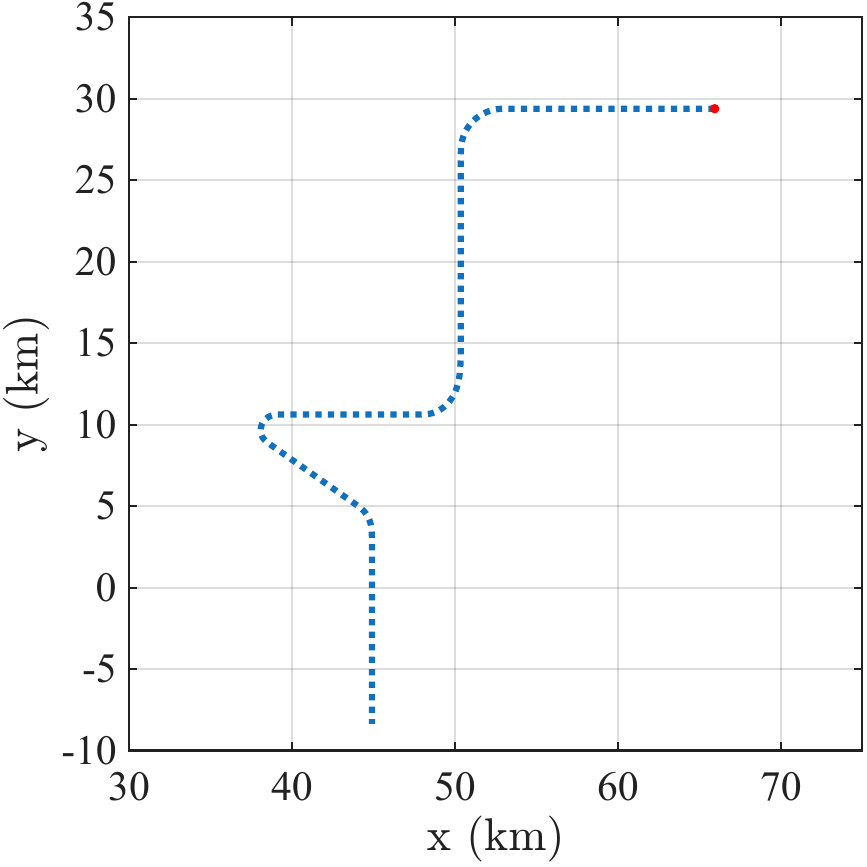}
  \caption{Fixed true target trajectory. The red dot indicates the start of the trajectory.}
  \label{fig:true_target_trajectory}
\end{figure}

We consider a 2D tracking problem with the deterministic target trajectory shown in Fig.~\ref{fig:true_target_trajectory}, which is the trajectory~6 from the benchmark scenarios in~\cite{BlairWKB1998}.
The deterministic state trajectory consists of the kinematic data of the target at time instants $k=0,\ldots,K$, where $K=3760$ for the considered problem with sampling period $T = 0.05$\,s.

\begin{figure}[!t]
  \centering
  \includegraphics[width=0.99\linewidth]{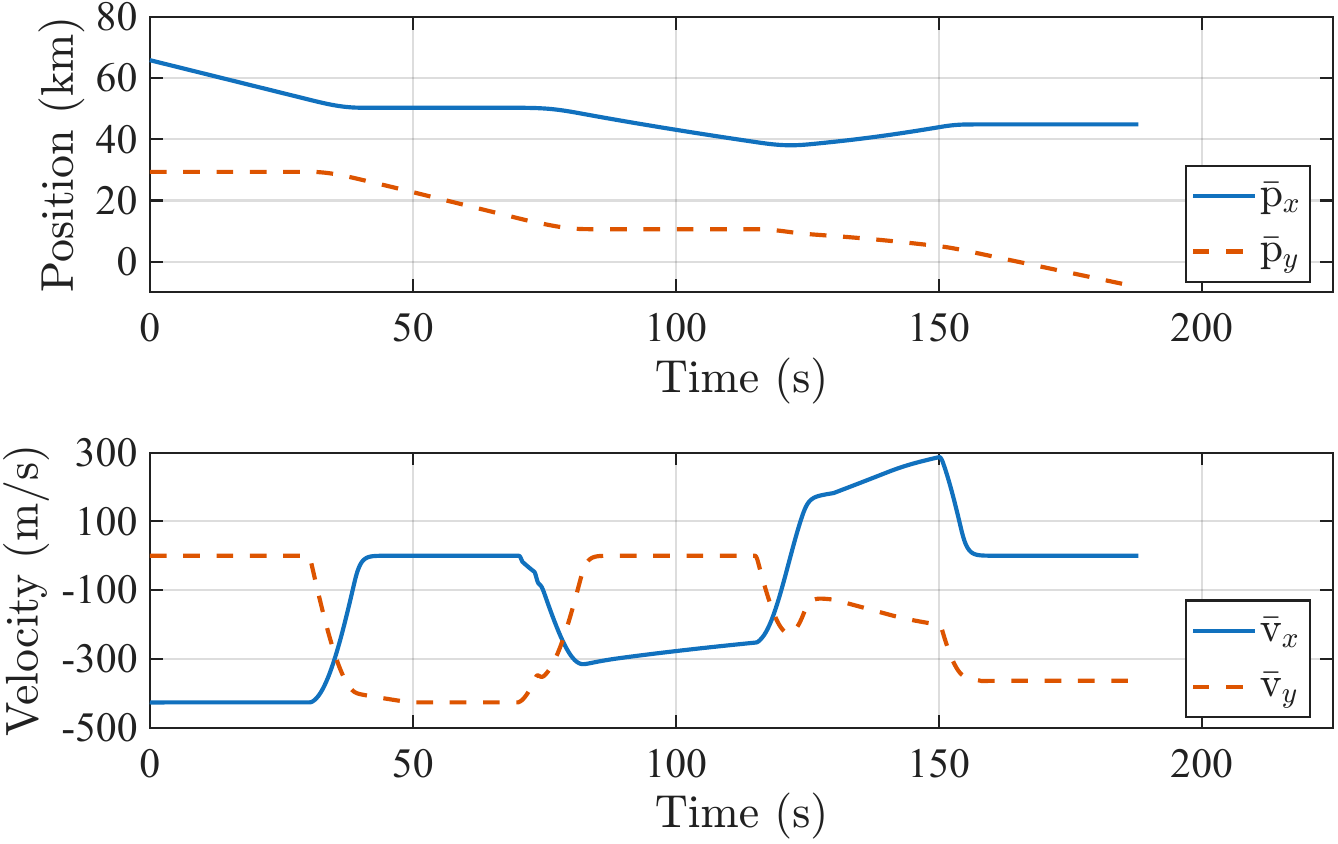}
  \caption{True target position and velocities.}
  \label{fig:true_target_positionandvelocities}
\end{figure}

 The true state vector $\xbbar_k\triangleq  \left[
\bar{\p}_{x,k},\,\bar{\p}_{y,k},\,\bar{\v}_{x,k},\,\bar{\v}_{y,k} \right]^\t$ is composed of true positions $\bar{\p}_{x,k}$, $\bar{\p}_{y,k}$, and true velocities $\bar{\v}_{x,k}$, $\bar{\v}_{y,k}$ illustrated in Fig.~\ref{fig:true_target_positionandvelocities}.  
The position measurements $\yb_k$, $k=0,\ldots,K$, are obtained by corrupting the true target positions $\bar{\p}_{x,k}$, $\bar{\p}_{y,k}$ of the target with zero-mean Gaussian measurement noise using~\eqref{eqn:true_measurement_model} with the following parameters.
\begin{align}
\Hbbar =& \left[\begin{array}{cc}
        \Ib_2 & \zerob_2
    \end{array}\right],& \Rbbar =&\,\mathrm{diag}
    \left(
        \bar{\sigma}_x^2, \,\bar{\sigma}_y^2
      \right),
\end{align} 
where $\bar{\sigma}^2_x=2000\,\mathrm{m}^2$ and $\bar{\sigma}^2_y=\,2000\,\mathrm{m}^2$ are the true measurement noise variances. 

The Kalman filter and smoother assume the standard nearly constant velocity state-space model with the state vector $\xb_k$ composed of the same elements as $\xbbar_k$ with $\xb_0\sim \N(\xb_0; \mub_0, \Sigmab_0)$, where 
\begin{subequations}
\begin{align}
\mub_0 =&\, \left [
    70000\,\mathrm{m}, \,25000\,\mathrm{m} ,\,-300\,\mathrm{m}/\mathrm{s},\,-100\,\mathrm{m}/\mathrm{s}\right ]^\t,
\\
\Sigmab_0 =&\, \mathrm{diag\left(
2000\,\mathrm{m}^2 ,\,2000\,\mathrm{m}^2 ,\,200\,\mathrm{m}^2/\mathrm{s}^2 ,\,200\,\mathrm{m}^2/\mathrm{s}^2
\right)}.
\end{align}
\end{subequations}
The parameters of the assumed state model~\eqref{eqn:assumed_state_equation} are given as
\begin{align}
\Fb=& \left[\begin{array}{cc} 
        \Ib_2 & T\,\Ib_2 \\
        \zerob_2 & \Ib_2
    \end{array}\right],&\Qb =&\, \sigma_a^2\left[\begin{array}{cc}
        \frac{T^3}{3}\,\Ib_2 &  \frac{T^2}{2}\, \Ib_2 \\
        \frac{T^2}{2} \,\Ib_2 & T\,\Ib_2
    \end{array}\right],
\end{align}
where $\sigma_a^2 = 10\,\mathrm{m}^2/\mathrm{s}^3 $. The parameters of the assumed measurement model~\eqref{eqn:assumed_measurement_model} are given as
\begin{align}
\Hb =&\,\eta\,\Hbbar =\eta\left[\begin{array}{cc}
        \Ib_2 & \zerob_2
    \end{array}\right],&\Rb =&\,\mathrm{diag}
    \left(\sigma_x^2 ,\,\sigma_y^2
    \right),
\end{align}
where $\eta = 0.99$ is a multiplicative bias parameter representing a mismatch between the true and assumed measurement models and $\sigma^2_x=\sigma^2_y=1800\,\mathrm{m}^2$.
The analytical MSE performance of the Kalman filter and smoother is obtained recursively using the proposed formulae given in Section~\ref{sec:AnalyticalRecursiveMSE}. These results are compared with the analytical methods derived in the previous study~\cite{KurtOrguner2025}, namely the horizon-recursive method. In addition, the results are compared with the numerical MSE values obtained by a Monte Carlo study. A total of $N_{\text{mc}}=10^5$ Monte Carlo runs were performed, each using a different realization of the measurement sequence $\yb_{0:K}$ from the true target trajectory via the model~\eqref{eqn:true_measurement_model} in each run. 
\begin{figure}[!t]
\vspace{3pt}
\centering
\includegraphics[width=0.99\linewidth]{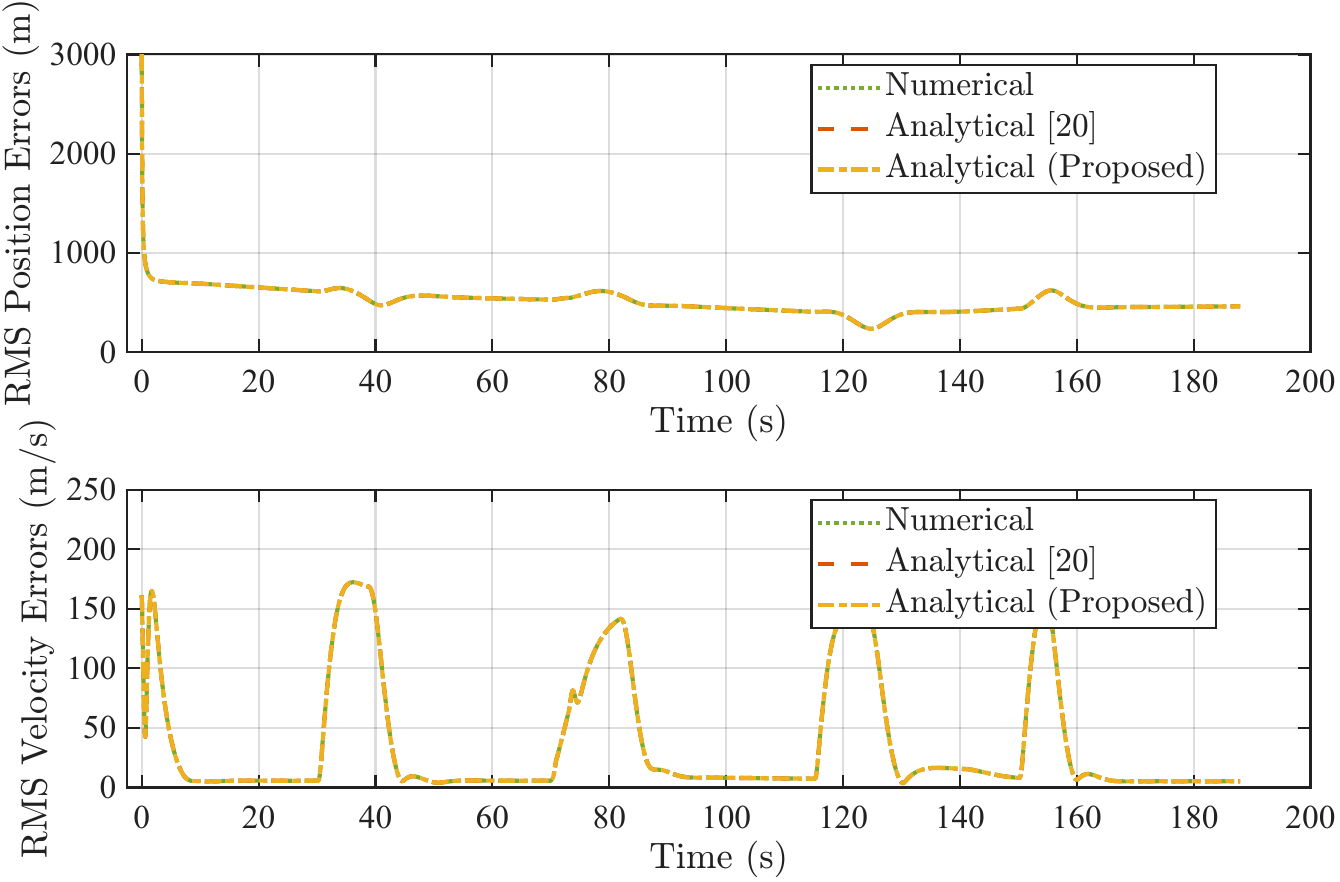}
\caption{Comparison of numerical and analytical RMS position and velocity errors for the Kalman filter.}
\label{fig:KF_RMSE}
\end{figure}
\begin{figure}[!t]
\centering
\includegraphics[width=0.99\linewidth]{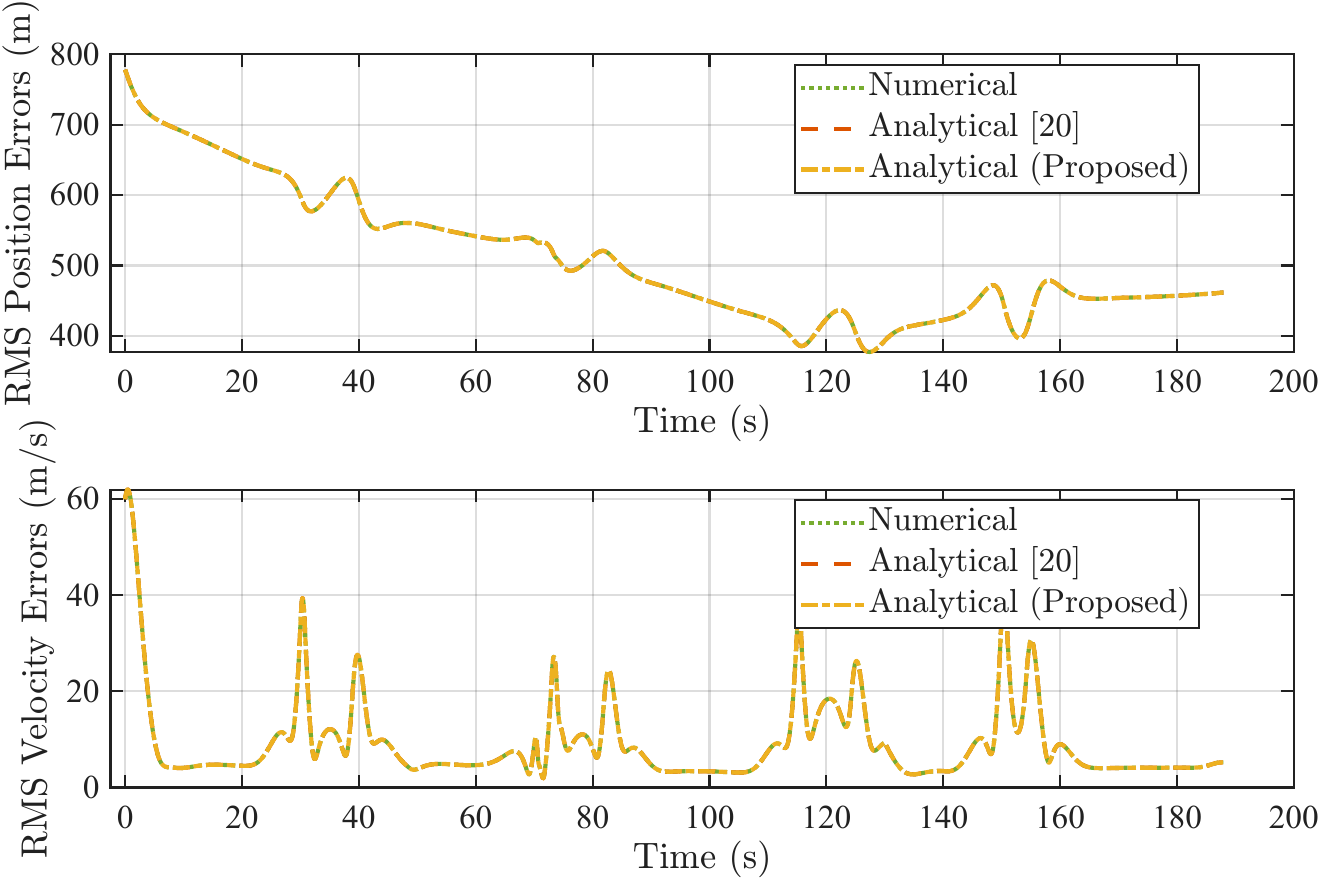}
\caption{Comparison of numerical and analytical RMS position and velocity errors for the Kalman smoother.}
\label{fig:KS_RMSE}
\end{figure}

For this specific problem, the execution time of each method is given in Table~\ref{tab:execution_times}. The aforementioned simulations were conducted using MATLAB R2025a on a MacBook Pro M4 device. 

\begin{table}[h!]
\centering
\caption{Execution Times of Compared Methods}
\begin{tabular}{c c}
\toprule
\textbf{Method} & \textbf{Execution Time (s)} \\
\midrule
Numerical              & 2532.0149 \\
Analytical~\cite{KurtOrguner2025}  & 60.6418 \\
Analytical (Proposed) & 0.1024 \\
\bottomrule
\end{tabular}\\
\label{tab:execution_times}
\end{table}

The numerical RMS position and velocity errors for the Kalman filter and smoother are shown in Figs.~\ref{fig:KF_RMSE} and~\ref{fig:KS_RMSE}, respectively, along with the analytical results.
The proposed method exhibits excellent agreement with both numerical results and previously established analytical formulations, while maintaining linear time complexity. Consequently, it is particularly well suited for benchmark studies involving long trajectories.

\section{Conclusions and Future Work}\label{sec:conclusions}
Efficient analytical recursive MSE expressions have been proposed for the Kalman filter and Kalman smoother under deterministic state trajectories and measurement model mismatch. The proposed method enables accurate performance prediction with linear time complexity and has been validated through simulation studies, demonstrating both accuracy and computational efficiency. An interesting future research direction would be to apply the proposed MSE framework to nonlinear systems under model mismatch.

\bibliographystyle{IEEEtran}
\bibliography{IEEEabrv, references}

@STRING{IEEE_J_AES        = "{IEEE} Trans. Aerosp. Electron. Syst."}

@STRING{IEEE_J_AC         = "{IEEE} Trans. Autom. Control"}

@article{Kalman1960,
    author = {Kalman, R. E.},
    title = {A New Approach to Linear Filtering and Prediction Problems},
    journal = {Journal of Basic Engineering},
    volume = {82},
    number = {1},
    pages = {35-45},
    year = {1960},
    month = mar,
}

@article{WallWS1981,
author = {Wall, Jr., Joseph E. and Willsky, Alan S. and Sandell, Jr., Nils R.},
title = {On the fixed-interval smoothing problem},
journal = {Stochastics},
volume = {5},
number = {1-2},
pages = {1--41},
year = {1981},
publisher = {Taylor \& Francis},
doi = {10.1080/17442508108833172},
}

@article{MAYNE1966,
title = {A solution of the smoothing problem for linear dynamic systems},
journal = {Automatica},
volume = {4},
number = {2},
pages = {73-92},
year = {1966},
author = {D.Q. Mayne}}

@phdthesis{Fraser1967,
  title={A new technique for the optimal smoothing of data},
  author={Fraser, Donald Charles},
  year={1967},
  school={Massachusetts Institute of Technology}
}

@article{RTS1965,
author = {Rauch, H.E. and Tung, F. and Striebel, C. T.},
title = {Maximum likelihood estimates of linear dynamic systems},
journal = {AIAA Journal},
volume = {3},
number = {8},
pages = {1445--1450},
year = {1965},
}

@INPROCEEDINGS{KurtOrguner2025,
  author={Kurt, Batın and Orguner, Umut},
  booktitle={International Conference on Information Fusion (FUSION)}, 
  title={{MSE} of {K}alman Filter and Smoother for Fixed Non-Random State Trajectories}, 
  year={2025},
  pages={1-8},
  }

@book{Thrun2005Robotics,
author = {Thrun, Sebastian and Burgard, Wolfram and Fox, Dieter},
title = {Probabilistic Robotics},
year = {2005},
publisher = {MIT Press},
address = {Cambridge, MA}
}

@book{Barfoot2024,
  author    = {Timothy D. Barfoot},
  title     = {State Estimation for Robotics},
  publisher = {Cambridge University Press},
  year      = {2024},
  address   = {Cambridge, United Kingdom},
}

@book{Kay_EstimationTheory,
	author = {Kay, S. M.},
	title = {Fundamentals of Statistical Signal Processing: Estimation Theory},
	year = {1993},
	publisher = {Prentice-Hall},
        address = {Upper Saddle River, NJ}
}

@ARTICLE{BlairWKB1998,
  author={Blair, W.D. and Watson, G.A. and Kirubarajan, T. and Bar-Shalom, Y.},
  journal=IEEE_J_AES, 
  title={Benchmark for radar allocation and tracking in {ECM}}, 
  year={1998},
  volume={34},
  number={4},
  pages={1097--1114},
  }

@book{Haykin2002Adaptive,
  address = {Upper Saddle River, NJ},
  author = {Haykin, Simon},
  edition = {4th},
  publisher = {Prentice Hall},
  title = {Adaptive Filter Theory},
  year = {2002}
}

@book{Groves2013,
  author={Groves, Paul},
  title={Principles of GNSS, Inertial, and Multisensor Integrated Navigation Systems},
  year={2013},
  publisher={Artech House},
  address={Norwood, MA},
}

@book{stengel1994optimal,
  title={Optimal Control and Estimation},
  author={Stengel, Robert F},
  year={1994},
  publisher={Courier Corporation}
}

@book{simon2006optimal,
  title={Optimal State Estimation: Kalman, H infinity, and Nonlinear Approaches},
  author={Simon, Dan},
  year={2006},
  publisher={John Wiley \& Sons},
address   = {Hoboken, NJ},
}

@book{bar2001estimation,
  title={Estimation with Applications to Tracking and Navigation: {T}heory Algorithms and Software},
  author={Bar-Shalom, Yaakov and Li, X Rong and Kirubarajan, Thiagalingam},
  year={2001},
  publisher={John Wiley \& Sons},
  address = {Hoboken, NJ}
}

@INPROCEEDINGS{TeichnerM2023KF,
  author={Teichner, Ron and Meir, Ron},
  booktitle={European Control Conference (ECC)}, 
  title={Discrete-Time {K}alman Filter Error Bounds in the Presence of Misspecified Measurements}, 
  year={2023},
  pages={1-7},
}

@ARTICLE{TeichnerM2023KS,
title = {Kalman smoother error bounds in the presence of misspecified measurements},
author={Teichner, Ron and Meir, Ron},
journal = {IFAC-PapersOnLine},
volume = {56},
number = {2},
pages = {10252--10257},
year = {2023},
note = {{IFAC} World Congress},
issn = {2405--8963}
}

@ARTICLE{GeSDW2016,
  author={Ge, Quanbo and Shao, Teng and Duan, Zhansheng and Wen, Chenglin},
  journal=IEEE_J_AC, 
  title={Performance Analysis of the {K}alman Filter With Mismatched Noise Covariances}, 
  year={2016},
  volume={61},
  number={12},
  pages={4014--4019},
  }

@ARTICLE{SangsukB1990,
  author={Sangsuk-Iam, S. and Bullock, T.E.},
  journal=IEEE_J_AC, 
  title={Analysis of discrete-time {K}alman filtering under incorrect noise covariances}, 
  year={1990},
  volume={35},
  number={12},
  pages={1304--1309},
  }

@INPROCEEDINGS{Fritsche2016,
  author={Fritsche, Carsten and Orguner, Umut and Gustafsson, Fredrik},
  booktitle={International Conference on Acoustics, Speech and Signal Processing (ICASSP)}, 
  title={On parametric lower bounds for discrete-time filtering}, 
  year={2016},
  pages={4338--4342},
  }

@ARTICLE{BlackmanDBP1999,
  author={Blackman, S.S. and Dempster, R.J. and Busch, M.T. and Popoli, R.F.},
  journal=IEEE_J_AES, 
  title={{IMM/MHT} solution to radar benchmark tracking problem}, 
  year={1999},
  volume={35},
  number={2},
  pages={730--738},
}


\end{document}